\documentclass[twocolumn,prl,prb,superscriptaddress]{revtex4-1}
\usepackage[latin9]{luainputenc}
\usepackage{amsmath}
\usepackage{graphicx}

\makeatletter

\providecommand{\tabularnewline}{\\}

\usepackage{pifont}
\usepackage{graphicx}
\usepackage{dcolumn}
\usepackage{bm}

\usepackage{hyperref}
\usepackage[caption=false,font=footnotesize]{subfig}

\usepackage{amsfonts}

\makeatother

\begin{document}

\title{Distributed Coupling Accelerator Structures: A New Paradigm for High Gradient Linacs}

\author{Sami Tantawi}

\affiliation{SLAC National Accelerator Laboratory, CA, USA}

\author{Mamdouh Nasr}

\thanks{Corresponding author: mamdouh@slac.stanford.edu}

\affiliation{SLAC National Accelerator Laboratory, CA, USA}

\author{Zenghai Li}

\affiliation{SLAC National Accelerator Laboratory, CA, USA}

\author{Cecile Limborg}

\affiliation{SLAC National Accelerator Laboratory, CA, USA}

\author{Philipp Borchard}

\affiliation{Dymenso LLC, San Francisco}
\begin{abstract}
\textbf{Abstract:} We present a topology for linear accelerators (linacs) that permits larger degrees of freedom for the optimization of individual cavity shapes. The power is distributed to the cavities through a waveguide with periodic apertures that guarantees the correct phases and amplitudes along the structure. This topology optimizes the power consumption and efficiency and allows the manipulation of the surface fields for high gradient operation. It also provides a possibility for low-temperature manufacturing techniques for use with novel materials. This greatly enhanced performance for both normal and superconducting linacs. We present a design and an experimental demonstration of this linac. 
\end{abstract}
\maketitle
A linear particle accelerator (linac) accelerates charged particles using oscillating electric field formed within RF cavities (cells) that are joined together to form a beamline. Charged particles gain energy as they travel along the beamline. Typically, RF power is fed to the linac from one point and flows through adjacent cells using coupling holes that also serves as a beam tunnel for charged particles. Consequently, the linac design process requires careful consideration of the coupling between adjacent cells. This limits the ability of designers to optimize the cell shape for efficiency and or gradient handling capability. \cite{karzmark1993medical,wangler2008rf,maury1999handbook_acc_eng}

The distributed coupling linac, presented here, have a topology that allows feeding each accelerator cell independently using a periodic feeding network. This adds another degree of freedom to the design; the coupling between cells can have a wider acceptable range including minimizing it to negligible values. This leads to huge flexibility in the geometry optimization of the accelerator cells which can be used to attain the highest possible shunt impedance for a set of cells or craft the field along the cell walls to allow for higher gradient operation. \cite{tantawi2016distributed} 

Any concept for a distributed feeding network is required to provide, typically, an equal amount of power to each identical cell and simultaneously provide the appropriate phased advance for each cell. For a typical electron Linac with particle moving at nearly the speed of light, the phase advance per cell is $2\pi P/\lambda$; where $P$ is the periodic separation between cells and $\lambda$ is the free space wavelength. With the exception of a dielectric-free coaxial line, this phase advance cannot be provided by any form of wave guiding structure which always have a guided wavelength $\lambda_{g}>\lambda$, the free space wavelength. Using a coaxial line to feed a set of cavities has been attempted before \cite{sundelin1977parallel}. This approach has not been applied since this initial publication; the use of coaxial line is truly impractical for many reasons including losses, high field handling capabilities, manufacturing difficulties, and inability to extend the concepts to higher frequencies. 

Our approach, on the other hand, uses waveguides oriented so that its center E-plane coincide with an E-plane of the accelerator cell\textquoteright s $\mathrm{TM_{010}}$ mode, which is any plane that contains the center axis of the accelerator cell. Because the guided wavelength does not match the free-space wavelength, one can imagine solutions where the distribution wave guide is bent like a serpentine to achieve the appropriate phase advance. This is valid, but one can use more than a single manifold. A natural interval for taping into the manifold, as would be seen below, is every $(m\lambda_{g})/2$; where $\lambda_{g}$ is the guided wavelength within the manifold, and m is an integer. Therefore, for a structure with a phase advance/cavity of $2\pi/n$ one can use $n$ manifolds each one of them is being taped every $\lambda_{g}/2$ with an interleaving $\pi$ phase delay incorporated at the cell feed lines; every cell is fed by a given manifold have the same phase. Since we have total freedom of choosing the phase advance between cells, the phase advance can be viewed as an optimization parameter. 

\begin{figure}
\begin{centering}
\includegraphics[width=8cm]{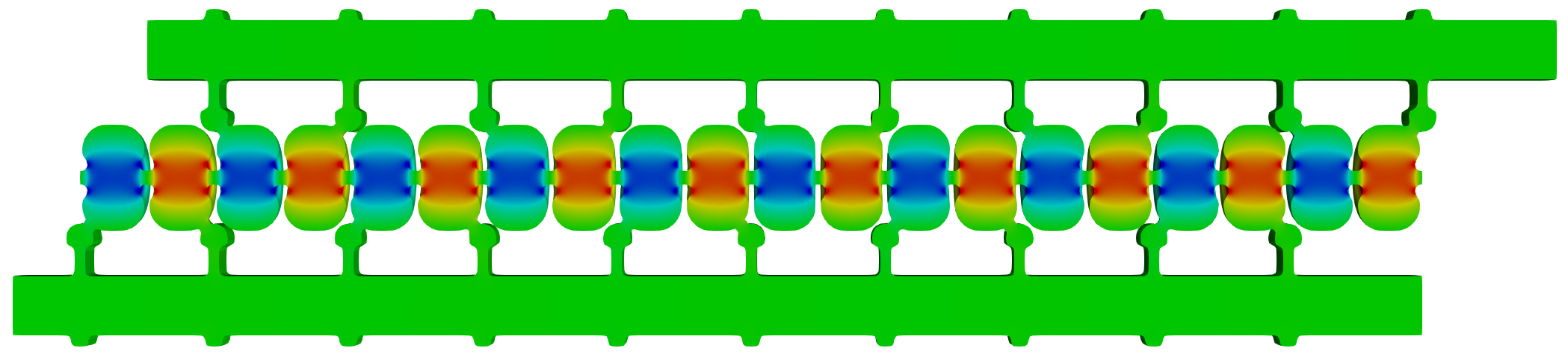}
\par\end{centering}
\begin{centering}
(a)
\par\end{centering}
\begin{centering}
\includegraphics[scale=0.25]{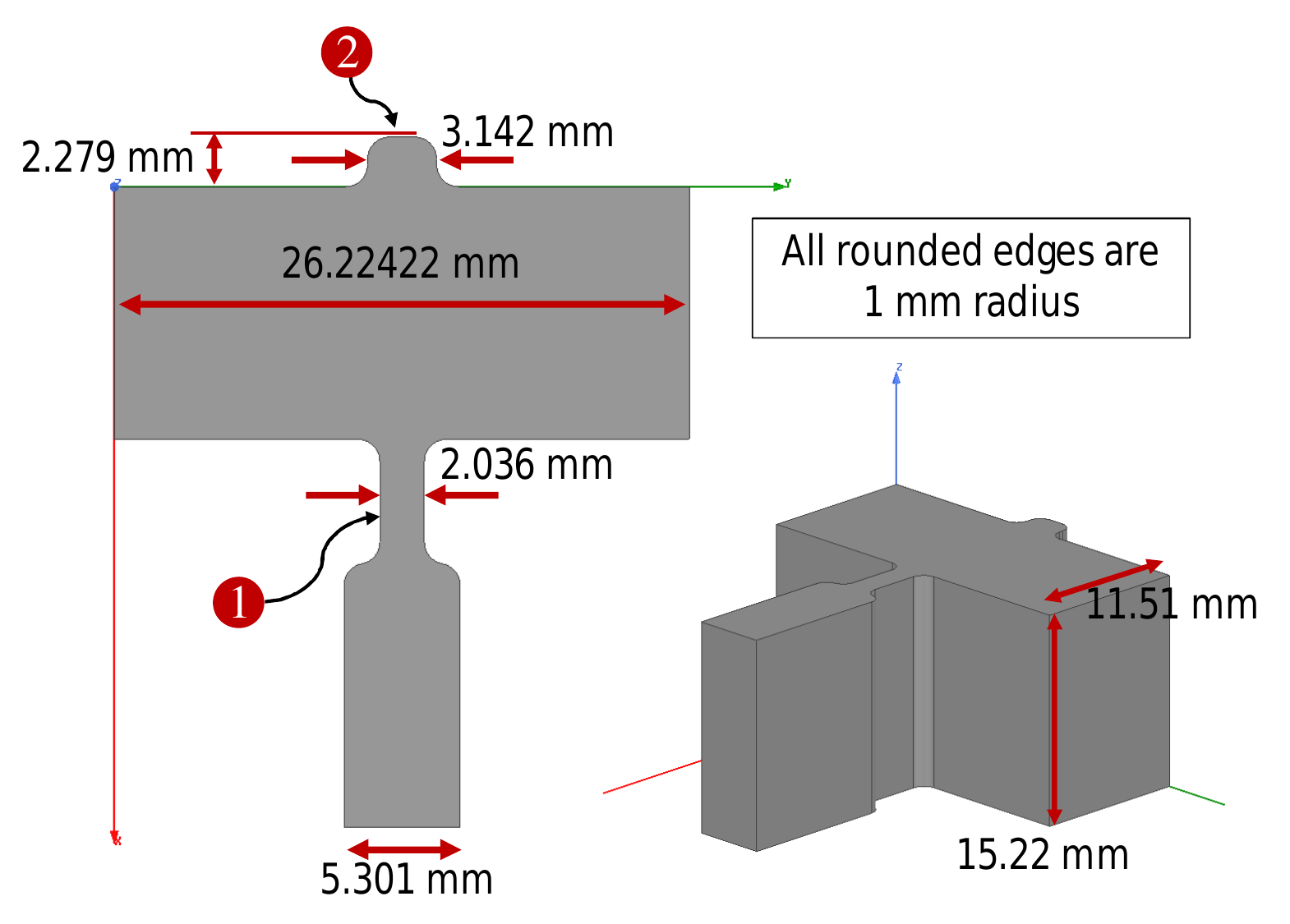}\includegraphics[scale=0.19]{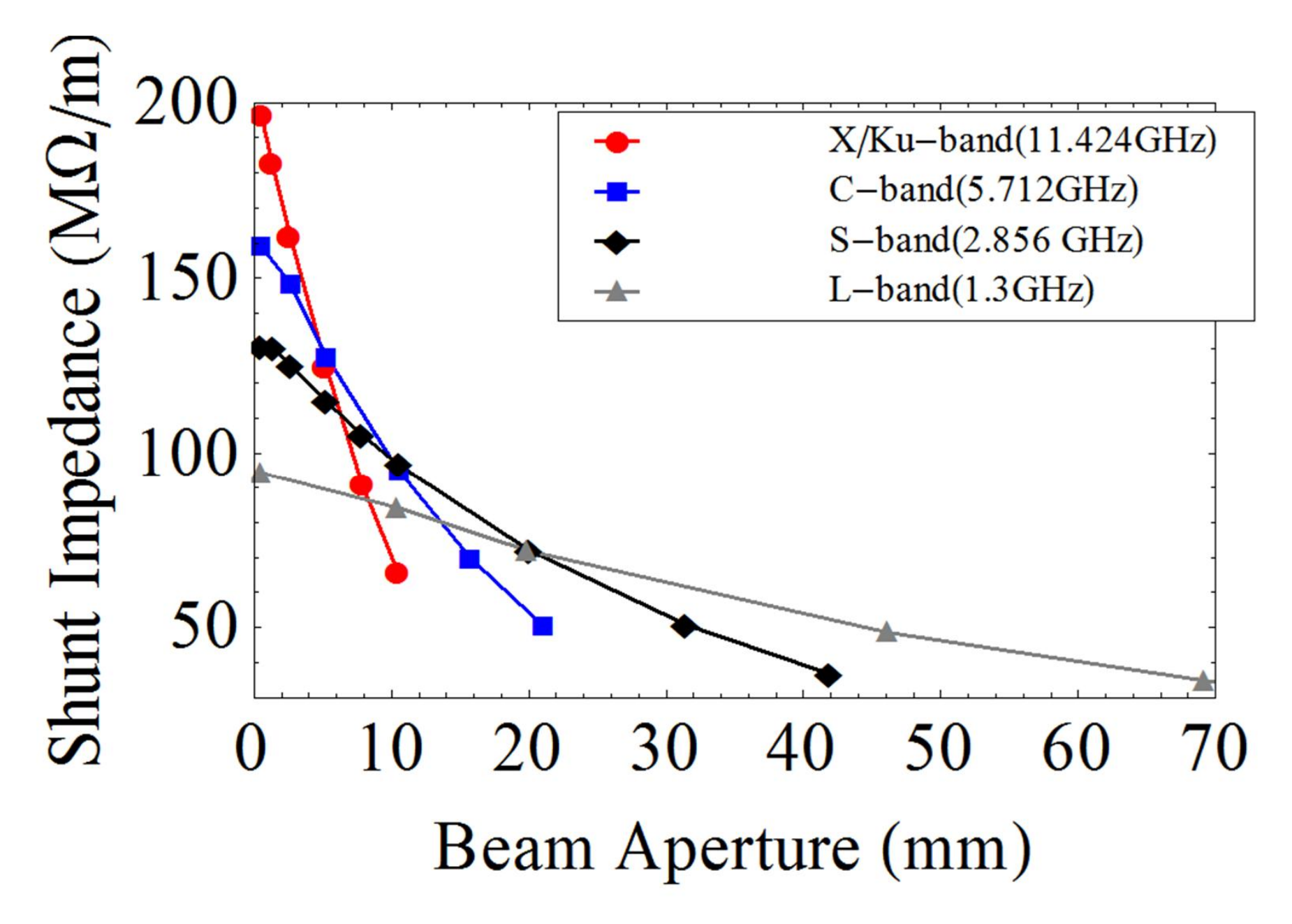}
\par\end{centering}
\begin{centering}
(b) $\hspace{3cm}$(c)
\par\end{centering}
\caption{\label{fig:Tjunction_Xband} (a) the full design of 20-cells $\pi$-mode distributed-coupling linac with identical cells at X-band (b) the T-junction design accelerator structure (N=10 cells per waveguide) that achieves the S-matrix for $\lambda_{g}/2$-periodicity, and (c) the shunt impedance for optimized designs utilizing the distributed-coupling technology for different aperture openings at different bands }
\end{figure}

A $\pi$ phase advance/cavity is a special case that simplify the design of the system. In this case, $n=2$, and hence, only two manifolds are needed. It turns out that this period is close to being optimal. For small beam apertures, a phase advance of $2\pi/3$ is slightly better in terms of shunt impedance \cite{miller1986comparison}. However, for simplicity, we made our initial design, shown in Fig. \ref{fig:Tjunction_Xband}(a), around the $\pi$ mode of operation. The manifold, in this case, can also be simplified by supplying the power from the manifold to one cavity every $\lambda_{g}/2$. Transverse-fields of the fundamental mode of a waveguide switch polarity every half guide wavelength. In this case, there is no need to modify the feeding line for every other cell as suggested before; feeding lines are the same for all the cavities. 

The manifold network will consist of a set of cascaded T-junctions. The power going through the tap-off ports to the cavities should achieve a minimal standing-wave ratio (SWR) with the feeding lines to the cavities; the feeding network, then, will have minimal influence on the cavity cells, resulting into two isolated systems (the feeding network and the cavity cells). To this end, the cavity coupling port should be designed to give a matched port when the cavity is loaded with the design beam current. 

The isolated manifold system, and with the assumption of a matched loads at the end of the tap-off ports, can be viewed as a transmission line with periodic loading and an open circuit termination at the end. With all the identical loads with impedance $Z_{l}$ located at $i\lambda_{g}/2$; where $i\in\mathbb{Z}$, the total load impedance can be shown with the help of elementary transmission line theory to be $Z_{l}/n$. Hence, the choice of $Z_{l}=nZ_{0}$; where $Z_{0}$ is the characteristic impedance of the line. With this choice, the line is segmented into a series of sections with the load impedance in the middle of the section and a quarter wave transmission line extending the section from either direction. This section can be viewed as a two-port network. Using scattering matrix representation which could be immediately deduced from the impedance matrix representation for each section \cite{montgomery1987principles}, we can write the scattering matrix for each section as 

\begin{equation}
S_{two}=\left(\begin{array}{cc}
\frac{-1}{2n+1} & \frac{1}{2n+1}-1\\
\frac{1}{2n+1}-1 & \frac{-1}{2n+1}
\end{array}\right)\label{eq:S_two}
\end{equation}

The scattering matrix representation, $S$, of the lossless three-port tap-off section can then be deduced. The 2x2 matrix entries of Eq. \eqref{eq:S_two} has to be the same as the upper left submatrix of $S$. $S$ has to be symmetrical because of reciprocity. Because it is loss-less, it satisfies the unitarity condition; $SS^{\dagger}=I$. The geometric symmetry of the system is expressed in this matrix if $s_{11}=s_{22}$ and $s_{13}=-s_{-32}$ ; the negative sign expresses the $\pi$ phase-shift across each section. Hence, 

\begin{equation}
S=\left(\begin{array}{ccc}
\frac{-1}{2n+1} & \frac{1}{2n+1}-1 & \frac{2i\sqrt{n}}{2n+1}\\
\frac{1}{2n+1}-1 & \frac{-1}{2n+1} & \frac{-2i\sqrt{n}}{2n+1}\\
\frac{2i\sqrt{n}}{2n+1} & \frac{-2i\sqrt{n}}{2n+1} & \frac{2n-1}{2n+1}
\end{array}\right)\label{eq:S_full}
\end{equation}

The next stage of the implantation requires the translation of this matrix into a realizable physical structure. This is done with the aid of the a more general scattering matrix representation of any three-port junction satisfying unitarity, reciprocity, and geometrical symmetry, but without satisfying the particular form of Eq. \ref{eq:S_full}. This matrix is derived in \cite{tantawi1997active_pulse_compression} and is given by 

\begin{equation}
S=\left(\begin{array}{ccc}
\frac{-\mathrm{e}^{i\phi}-\cos(\theta)}{2}\mathrm{e}^{i2\psi} & \frac{-\mathrm{e}^{i\phi}+\cos(\theta)}{2}\mathrm{e}^{i2\psi} & \frac{\sin(\theta)}{\sqrt{2}}\mathrm{e}^{i\vartheta}\mathrm{e}^{i\psi}\\
\frac{-\mathrm{e}^{i\phi}+\cos(\theta)}{2}\mathrm{e}^{i2\psi} & \frac{-\mathrm{e}^{i\phi}-\cos(\theta)}{2}\mathrm{e}^{i2\psi} & \frac{-\sin(\theta)}{\sqrt{2}}\mathrm{e}^{i\vartheta}\mathrm{e}^{i\psi}\\
\frac{\sin(\theta)}{\sqrt{2}}\mathrm{e}^{i\vartheta}\mathrm{e}^{i\psi} & \frac{-\sin(\theta)}{\sqrt{2}}\mathrm{e}^{i\vartheta}\mathrm{e}^{i\psi} & \cos(\theta)\mathrm{e}^{i2\vartheta}
\end{array}\right)\label{eq:S_general}
\end{equation}

After imposing reciprocity, unitarity and geometrical symmetry conditions, there are only 4 degrees of freedoms left, represented by $\theta$,$\phi$,$\psi$, and $\vartheta$. Hence, the physical structure shown in Fig. \ref{fig:Tjunction_Xband}(b), for the T-junction, with features labeled \ding{182} and \ding{183} have 4-dimensional degrees of freedom in terms of the depth and width of each feature. The first feature provides a control to achieve the design value for $|s_{33}|$ or $\theta$ and $\vartheta$, while the second one controls $\phi$, and $\psi$. Performing a numerical search through a finite element code quickly determines the correct dimensions for this junction so that it satisfy Eq. (\ref{eq:S_full}).

As described, the practical implementation of a distributed feeding network provides independent power feeding to each accelerator cell, and thus enables designing for completely isolated cells. In this case, the accelerator design converges to a single-cell design with identical accelerator cells. This simplicity in the design concept opens the door to investigate new accelerator-cell geometries. We implemented new optimization approaches that utilize circular, elliptical, or even more general shapes using Splines to provide optimum cell designs with much-enhanced accelerating parameters \cite{nasr2018new_geom}. 

Figure \ref{fig:Tjunction_Xband}(c) shows the shunt impedance for optimized designs utilizing the distributed-coupling technology for different aperture openings at different bands. This shows improved shunt impedance for all bands. Consequently, the distributed coupling linac provides a scalable technology with enhanced shunt impedance that enables reaching high duty factors compared to any conventional particle accelerator.

We applied the distributed feeding scheme to a $\pi$-mode standing-wave linac design operating at X-band (11.424 GHz). The linac consists of 20 accelerating cells; i.e., N=10 cells per manifold; the iris diameter is 2.6 mm. As described before, the S-matrix in Eq. \eqref{eq:S_full} for N=10 was achieved by introducing two features to the T-junction as shown in Fig. \ref{fig:Tjunction_Xband}(b). 

\begin{figure}
\begin{centering}
\includegraphics[width=7cm]{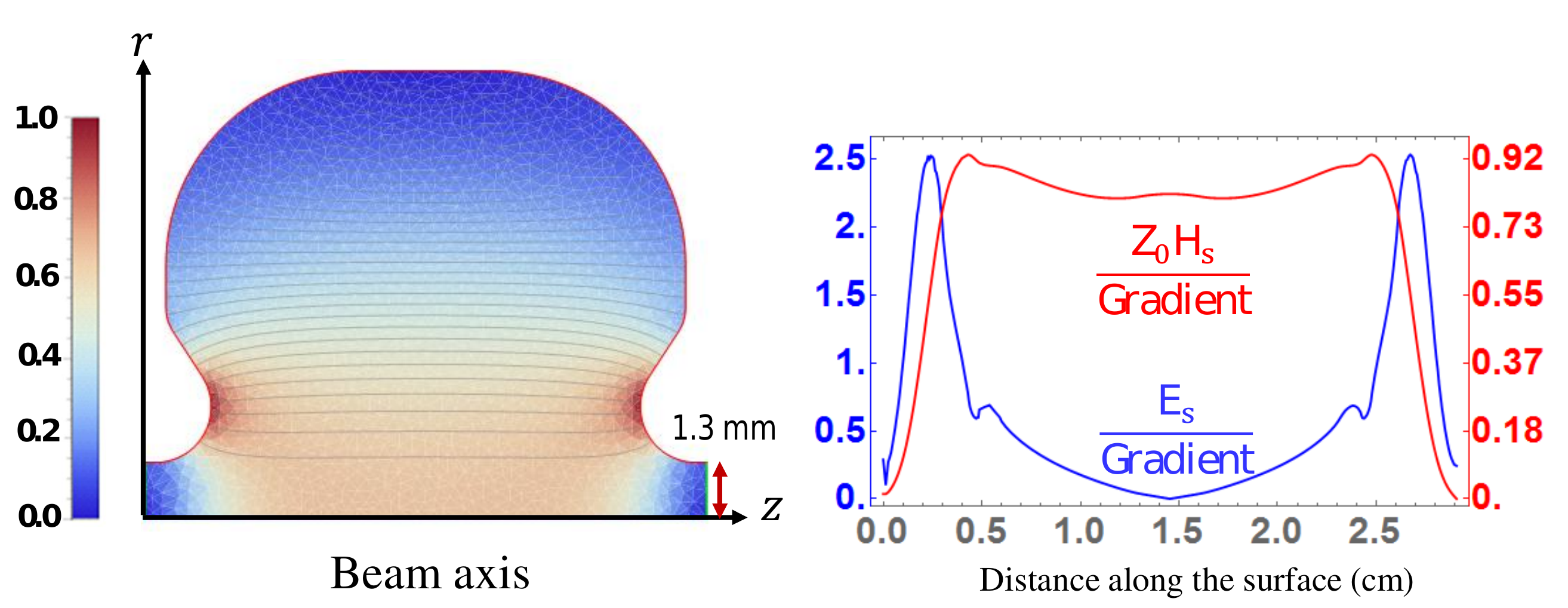}
\par\end{centering}
\begin{centering}
(a) $\hspace{75bp}$(b)
\par\end{centering}
\caption{\label{fig:cell_shape_and_fields} (a) the optimized cell-design for X-band linac with iris-diameter of 2.6 mm with the electric field plot, and (b) a plot of the surface electric and magnetic fields for the accelerator cell.}
\end{figure}

In our design, we optimized for minimal peak magnetic field on the surface of the accelerator cells. Many studies showed the correlation between breakdown rates in accelerator structures and the peak magnetic fields on the accelerator walls \cite{Exp_study_RF_pulsed_heating,HighGradAcc_pulsehheating,pritzkau1999possible_pulsed_heating,pritzkau2001rf}. Consequently, the optimized design, shown in Fig. \ref{fig:cell_shape_and_fields}(a), achieves a superior shunt-impedance of 160 $\mathrm{M\Omega/m}$ at peak surface electric field to gradient ratio of 2.5. The full linac design that combines the feeding network with the accelerator cells is shown in Fig. \eqref{fig:Tjunction_Xband}(a) and a summary of the linac parameters is presented in Table \ref{tab:Summary-of-the}.

\begin{table}
\caption{\label{tab:Summary-of-the}Summary of the cavity accelerating parameters. The peak fields are for average gradient of 100 MV/m.}

\begin{ruledtabular} %
\begin{tabular}{ll}
Frequency  & 11.424 GHz\tabularnewline
Iris diameter & 2.6 mm\tabularnewline
Shunt impedance & 160 M$\Omega$/m\tabularnewline
Peak surface electric field & 250 MV/m\tabularnewline
Peak surface magnetic field & 0.25 MA/m\tabularnewline
$\mathrm{S_{c}}$ \cite{sc} & 7.36 W/$\mu\mathrm{m^{2}}$\tabularnewline
\end{tabular}\end{ruledtabular}
\end{table}

\begin{figure}
\includegraphics[width=8.5cm]{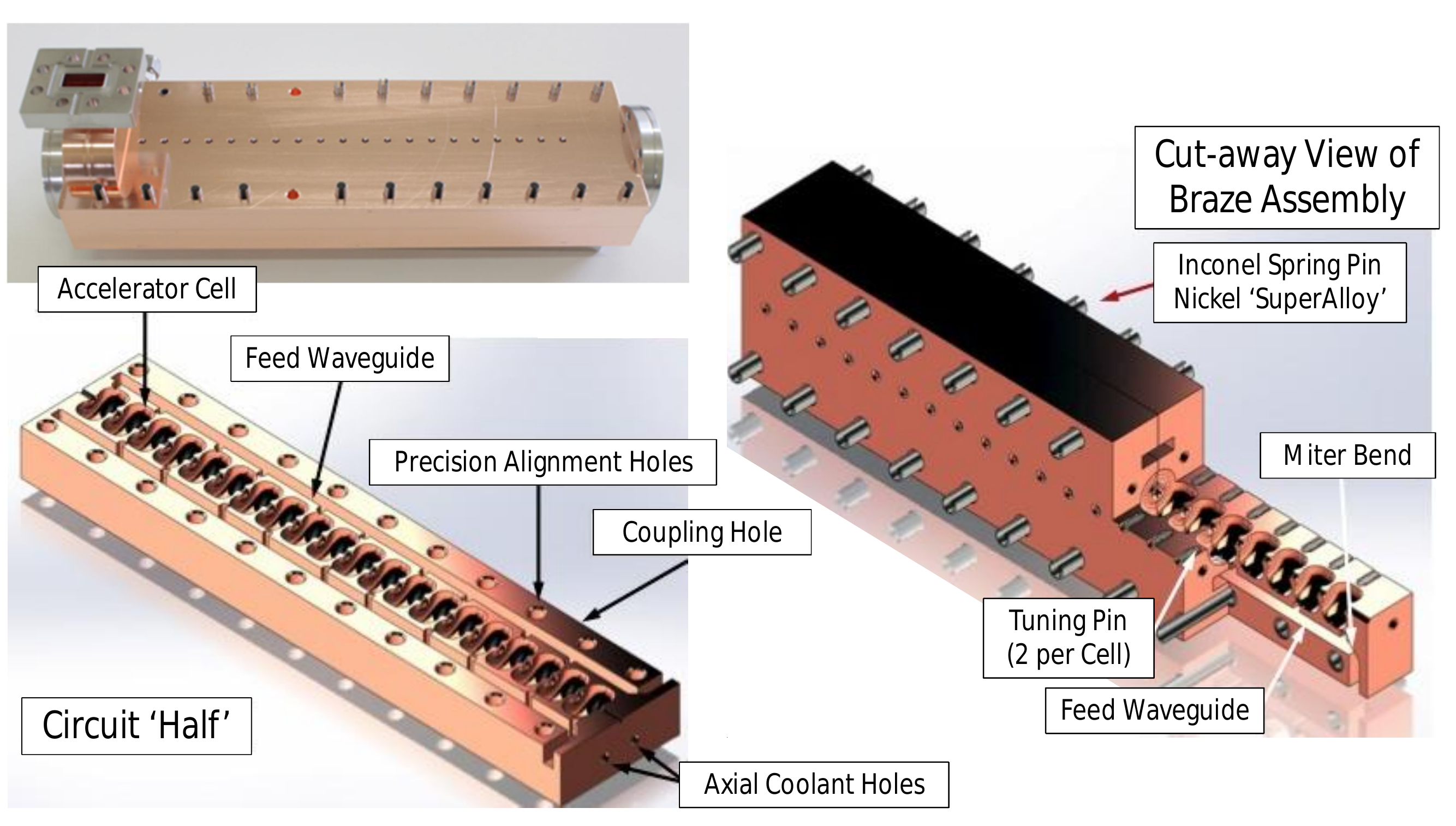}

\caption{\label{fig:acc_man_halfs}Illustration of the manufacturing of the distributed-coupling linac from two quasi-identical copper blocks including the tuning pins, alignment holes and cooling channels.}
\end{figure}

Each segment of the distributed-coupling accelerator structure can be manufactured from two blocks as shown in Fig. \ref{fig:acc_man_halfs} \cite{tantawi2016distributed}. Both the manifolds and the cavities have no currents crossing the plane which splits the structure in half along the long dimension of the manifold cross section. Manufacturing from two blocks reduces the complexity of manufacturing the structure and provides logical places for both the cooling tubes and the tuning holes. It also provides a possibility for low-temperature manufacturing techniques including electron beam welding, and thereby allowing the use of un-annealled copper alloys, and hence greater tolerance to high gradient operation. The circuit halves are aligned with an elastic averaging technique \cite{willoughby2005elastically_avg,borchard2018fabrication}.

\begin{figure}
\includegraphics[width=8.5cm]{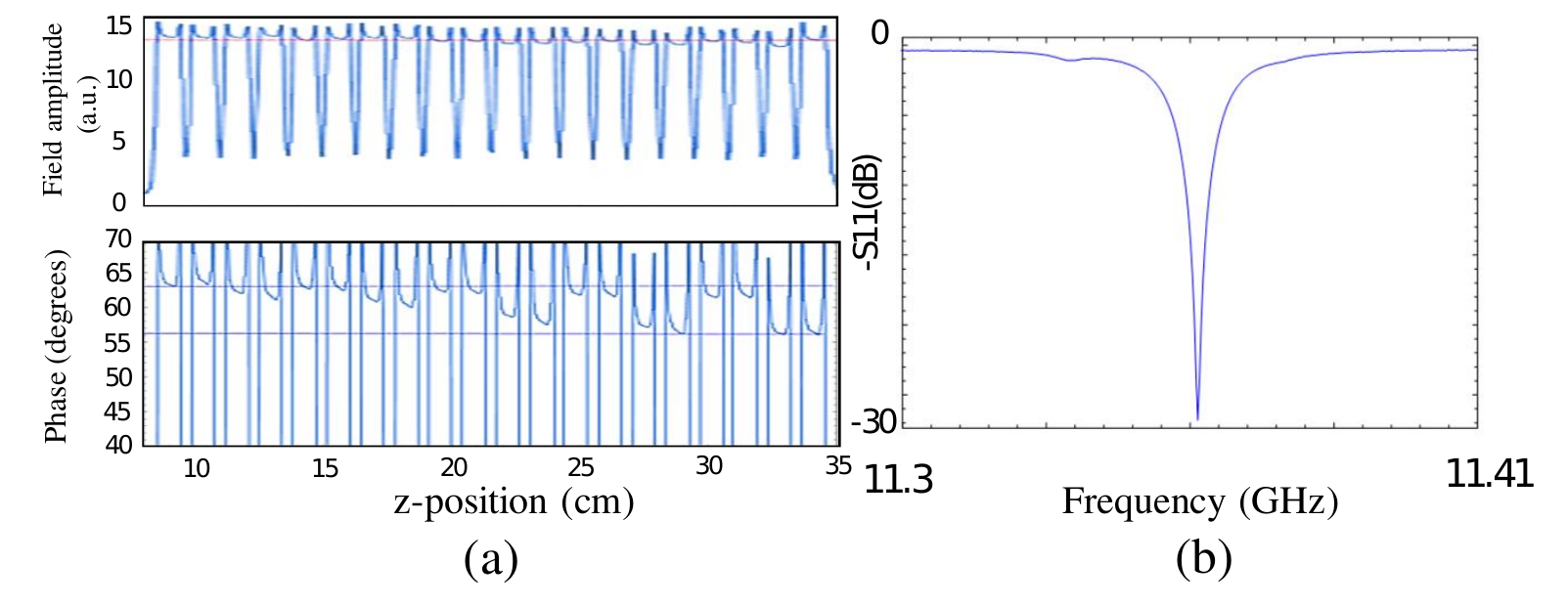}

\caption{\label{fig:cold_test_data}The cold test data for the manufactured 20-cells X-band standing-wave distributed-coupling linac showing (a) the amplitude and phase along the structure as well as (b) the frequency spectrum. The structure features a single resonance frequency rather than 20 resonances for conventional coupled structures. }
\end{figure}

The accelerator structure was manufactured and cold-tested before high power operation. Tuning was accomplished by shorting all the cells, with a conductive rod, except the one cell being tuned. The resultant structure, as expected, had a single resonance frequency rather 20 resonances for conventional structures with 20 coupled-accelerator cells \cite{wangler2008rf}. Figure \ref{fig:cold_test_data} shows the frequency response data together with the bead pull \cite{steele1966nonresonant} data that characterized for the amplitude and phase along the structure.

\begin{figure}
\includegraphics[width=8.5cm]{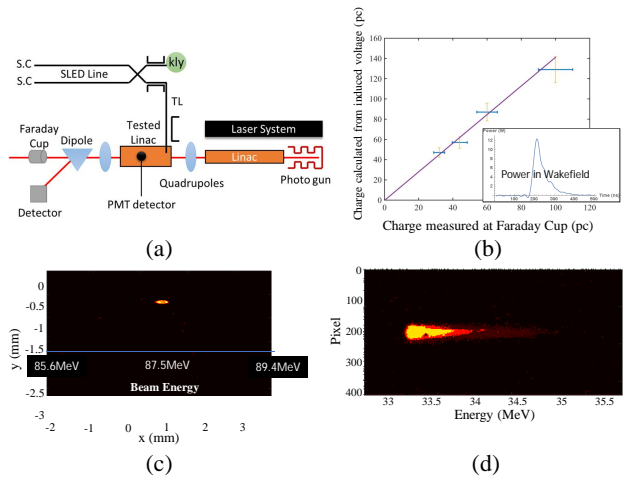}

\caption{\label{fig:grad_wake_curr}(a) The experimental setup for the distributed coupling linac testing (b) the measured Charge with Faraday Cup and Calculated from Induced Wakefield, (c) the beam shape and energy measurements at the output of the distributed-coupling linac for entrance energy of 63.5 MeV/m showing a gain of 24 MeV at gradient of 100 MV/m, and (d) the peak dark-current energy at gradient of 140 MV/m.}
\end{figure}

The structure is tested for high power operation at SLAC National Accelerator Laboratory with the experimental setup shown in Fig. \ref{fig:grad_wake_curr}(a). The electron beam is generated using a photo-gun with a Laser control system to generate and shape the electron beam. The electron bunches are then accelerated using an NLC-type Linac T0105 \cite{wang2004accelerator} and enters the distributed-coupling linac at 63.5 MeV. The distributed-coupling linac is fed from a single X-band klystron followed by a pulse compressor (the multimoded SLED-II \cite{Tantawi2005}), which was installed after the structure showed successful demonstration of ultra-high-gradient capability and the gradient was limited by the available klystron power. Along the beamline, there is a number of quadrupoles for beam focusing. Also, a dipole magnet is placed after the tested linac for beam-energy measurements. This is done by measuring the coil-current in the magnet that is needed to steer the beam to a fixed angle where the phosphorus detector is placed. Otherwise, when the dipole magnet is turned off, the beam go straight to a Faraday Cup to measure electron-bunch and dark-current charges . Finally, a photo-multiplayer tube (PMT) is placed on top of the structure to measure the bremsstrahlung radiation penetrating through the copper; this can be used for dark-current characterization and breakdown detection.

Before installation of the SLED-II pulse compressor, operation at 100 MV/m gradient with 16.5 MW of input power using 300ns square pulse was confirmed by measuring an energy gain of 24 MeV as shown in Fig. \ref{fig:grad_wake_curr}(c). Then, the RF shunt impedance was confirmed by measuring the induced Wakefield energy/bunch when an electron bunch passes through the structure. We calculated the value of the total charge from the cavity model. This calculated value is compared with the measured total charge at the Faraday Cup as shown in Fig. \ref{fig:grad_wake_curr}(b) shows good agreement between the two calculations, which validates the simulated shunt impedance value.

After the installation of the SLED-II RF pulse compressor, we were able to reach ultra-high gradients as high as 140 MV/m with no vacuum faults encountered in the structure. At high gradient levels we started to see an easily-measurable dark-current signals. As the charges migrate from the copper surface of the linac, some of them get captured on-axis and get accelerated along the beamline. This results in an energy spectrum with a peak value of the full acceleration gain. The peak dark-current energy was verified to be 35.7 MeV at a gradient of 140 MV/m as shown in Fig. \ref{fig:grad_wake_curr}(d). 

In summary, we developed a new linac architecture that feeds each accelerator cell independently. The distributed-coupling linac overcomes the limitations imposed on the cell shape to satisfy cell-to-cell coupling conditions; it hence allows for flexible geometry optimization of the accelerator cells. Consequently, the resultant designs achieve minimized surface fields, enabling high gradient operation, and maximized shunt impedance, enabling efficient operation. We also developed a new manufacturing technique that uses only two copper blocks to build the new linac architecture, enabling economical manufacturing. We applied the distributed coupling linac concept to a 20-cells X-band design and achieved a superior shunt impedance of 160 $\mathrm{M\Omega/m}$. We tested the linac for high power operation and verified the designed shunt impedance using different experimental measurements. We achieved a record gradient of 140 MV/m in our experiments. The distributed coupling linac topology opens the door for many applications and design spaces that include multi-frequency operation \cite{nasr2017novel,nasr2018design_dual,tantawi2016distributed}, and can be extended to superconducting accelerators \cite{welander2018parallel}.

\bibliographystyle{apsrev4-1}
\bibliography{PRLDist}

\begin{thebibliography}{22}%
\makeatletter
\providecommand \@ifxundefined [1]{%
 \@ifx{#1\undefined}
}%
\providecommand \@ifnum [1]{%
 \ifnum #1\expandafter \@firstoftwo
 \else \expandafter \@secondoftwo
 \fi
}%
\providecommand \@ifx [1]{%
 \ifx #1\expandafter \@firstoftwo
 \else \expandafter \@secondoftwo
 \fi
}%
\providecommand \natexlab [1]{#1}%
\providecommand \enquote  [1]{``#1''}%
\providecommand \bibnamefont  [1]{#1}%
\providecommand \bibfnamefont [1]{#1}%
\providecommand \citenamefont [1]{#1}%
\providecommand \href@noop [0]{\@secondoftwo}%
\providecommand \href [0]{\begingroup \@sanitize@url \@href}%
\providecommand \@href[1]{\@@startlink{#1}\@@href}%
\providecommand \@@href[1]{\endgroup#1\@@endlink}%
\providecommand \@sanitize@url [0]{\catcode `\\12\catcode `\$12\catcode
  `\&12\catcode `\#12\catcode `\^12\catcode `\_12\catcode `\%12\relax}%
\providecommand \@@startlink[1]{}%
\providecommand \@@endlink[0]{}%
\providecommand \url  [0]{\begingroup\@sanitize@url \@url }%
\providecommand \@url [1]{\endgroup\@href {#1}{\urlprefix }}%
\providecommand \urlprefix  [0]{URL }%
\providecommand \Eprint [0]{\href }%
\providecommand \doibase [0]{http://dx.doi.org/}%
\providecommand \selectlanguage [0]{\@gobble}%
\providecommand \bibinfo  [0]{\@secondoftwo}%
\providecommand \bibfield  [0]{\@secondoftwo}%
\providecommand \translation [1]{[#1]}%
\providecommand \BibitemOpen [0]{}%
\providecommand \bibitemStop [0]{}%
\providecommand \bibitemNoStop [0]{.\EOS\space}%
\providecommand \EOS [0]{\spacefactor3000\relax}%
\providecommand \BibitemShut  [1]{\csname bibitem#1\endcsname}%
\let\auto@bib@innerbib\@empty
\bibitem [{\citenamefont {Karzmark}\ \emph {et~al.}(1993)\citenamefont
  {Karzmark}, \citenamefont {Nunan},\ and\ \citenamefont
  {Tanabe}}]{karzmark1993medical}%
  \BibitemOpen
  \bibfield  {author} {\bibinfo {author} {\bibfnamefont {C.}~\bibnamefont
  {Karzmark}}, \bibinfo {author} {\bibfnamefont {C.}~\bibnamefont {Nunan}}, \
  and\ \bibinfo {author} {\bibfnamefont {E.}~\bibnamefont {Tanabe}},\ }\enquote
  {\bibinfo {title} {Medical electron accelerators},}\ \ (\bibinfo  {publisher}
  {McGraw-Hill, Incorporated, Health Professions Division},\ \bibinfo {year}
  {1993})\ Chap.\ \bibinfo {chapter} {3: Microwave Principles for
  Linacs}\BibitemShut {NoStop}%
\bibitem [{\citenamefont {Wangler}(2008)}]{wangler2008rf}%
  \BibitemOpen
  \bibfield  {author} {\bibinfo {author} {\bibfnamefont {T.}~\bibnamefont
  {Wangler}},\ }\enquote {\bibinfo {title} {Rf linear accelerators},}\ \
  (\bibinfo  {publisher} {Wiley},\ \bibinfo {year} {2008})\ Chap.\ \bibinfo
  {chapter} {4: Standard Linac Structures}\BibitemShut {NoStop}%
\bibitem [{\citenamefont {Maury}\ and\ \citenamefont
  {Wu}(1999)}]{maury1999handbook_acc_eng}%
  \BibitemOpen
  \bibfield  {author} {\bibinfo {author} {\bibfnamefont {T.}~\bibnamefont
  {Maury}}\ and\ \bibinfo {author} {\bibfnamefont {C.}~\bibnamefont {Wu}},\
  }\href {https://books.google.com/books?id=TILVCgAAQBAJ} {\emph {\bibinfo
  {title} {Handbook Of Accelerator Physics And Engineering (3rd Printing)}}}\
  (\bibinfo  {publisher} {World Scientific Publishing Company},\ \bibinfo
  {year} {1999})\BibitemShut {NoStop}%
\bibitem [{\citenamefont {Tantawi}\ \emph {et~al.}(2016)\citenamefont
  {Tantawi}, \citenamefont {Li},\ and\ \citenamefont
  {Borchard}}]{tantawi2016distributed}%
  \BibitemOpen
  \bibfield  {author} {\bibinfo {author} {\bibfnamefont {S.~G.}\ \bibnamefont
  {Tantawi}}, \bibinfo {author} {\bibfnamefont {Z.}~\bibnamefont {Li}}, \ and\
  \bibinfo {author} {\bibfnamefont {P.}~\bibnamefont {Borchard}},\ }\href@noop
  {} {\enquote {\bibinfo {title} {Distributed coupling and multi-frequency
  microwave accelerators},}\ } (\bibinfo {year} {2016}),\ \bibinfo {note} {uS
  Patent 9,386,682}\BibitemShut {NoStop}%
\bibitem [{\citenamefont {Sundelin}\ \emph {et~al.}(1977)\citenamefont
  {Sundelin}, \citenamefont {Kirchgessner},\ and\ \citenamefont
  {Tigner}}]{sundelin1977parallel}%
  \BibitemOpen
  \bibfield  {author} {\bibinfo {author} {\bibfnamefont {R.}~\bibnamefont
  {Sundelin}}, \bibinfo {author} {\bibfnamefont {J.}~\bibnamefont
  {Kirchgessner}}, \ and\ \bibinfo {author} {\bibfnamefont {M.}~\bibnamefont
  {Tigner}},\ }\href@noop {} {\bibfield  {journal} {\bibinfo  {journal} {IEEE
  Transactions on Nuclear Science}\ }\textbf {\bibinfo {volume} {24}},\
  \bibinfo {pages} {1686} (\bibinfo {year} {1977})}\BibitemShut {NoStop}%
\bibitem [{\citenamefont {Miller}(1986)}]{miller1986comparison}%
  \BibitemOpen
  \bibfield  {author} {\bibinfo {author} {\bibfnamefont {R.~H.}\ \bibnamefont
  {Miller}},\ }\href@noop {} {\emph {\bibinfo {title} {Comparison of standing
  wave and traveling wave structures}}},\ \bibinfo {type} {Tech. Rep.}\
  (\bibinfo {year} {1986})\BibitemShut {NoStop}%
\bibitem [{\citenamefont {Montgomery}\ \emph {et~al.}(1987)\citenamefont
  {Montgomery}, \citenamefont {Dicke},\ and\ \citenamefont
  {Purcell}}]{montgomery1987principles}%
  \BibitemOpen
  \bibfield  {author} {\bibinfo {author} {\bibfnamefont {C.~G.}\ \bibnamefont
  {Montgomery}}, \bibinfo {author} {\bibfnamefont {R.~H.}\ \bibnamefont
  {Dicke}}, \ and\ \bibinfo {author} {\bibfnamefont {E.~M.}\ \bibnamefont
  {Purcell}},\ }\href@noop {} {\emph {\bibinfo {title} {Principles of microwave
  circuits}}},\ \bibinfo {number} {25}\ (\bibinfo  {publisher} {Iet},\ \bibinfo
  {year} {1987})\BibitemShut {NoStop}%
\bibitem [{\citenamefont {Tantawi}\ \emph {et~al.}(1997)\citenamefont
  {Tantawi}, \citenamefont {Ruth}, \citenamefont {Vlieks},\ and\ \citenamefont
  {Zolotorev}}]{tantawi1997active_pulse_compression}%
  \BibitemOpen
  \bibfield  {author} {\bibinfo {author} {\bibfnamefont {S.~G.}\ \bibnamefont
  {Tantawi}}, \bibinfo {author} {\bibfnamefont {R.~D.}\ \bibnamefont {Ruth}},
  \bibinfo {author} {\bibfnamefont {A.~E.}\ \bibnamefont {Vlieks}}, \ and\
  \bibinfo {author} {\bibfnamefont {M.}~\bibnamefont {Zolotorev}},\ }in\
  \href@noop {} {\emph {\bibinfo {booktitle} {AIP Conference Proceedings}}},\
  Vol.\ \bibinfo {volume} {398}\ (\bibinfo {organization} {AIP},\ \bibinfo
  {year} {1997})\ pp.\ \bibinfo {pages} {813--821}\BibitemShut {NoStop}%
\bibitem [{\citenamefont {Nasr}\ and\ \citenamefont
  {Tantawi}(2018{\natexlab{a}})}]{nasr2018new_geom}%
  \BibitemOpen
  \bibfield  {author} {\bibinfo {author} {\bibfnamefont {M.}~\bibnamefont
  {Nasr}}\ and\ \bibinfo {author} {\bibfnamefont {S.}~\bibnamefont {Tantawi}},\
  }in\ \href@noop {} {\emph {\bibinfo {booktitle} {9th Int. Particle
  Accelerator Conf.(IPAC'18), Vancouver, BC, Canada, April 29-May 4, 2018}}}\
  (\bibinfo {organization} {JACOW Publishing, Geneva, Switzerland},\ \bibinfo
  {year} {2018})\ pp.\ \bibinfo {pages} {4395--4397}\BibitemShut {NoStop}%
\bibitem [{\citenamefont {Laurent}\ \emph {et~al.}(2011)\citenamefont
  {Laurent}, \citenamefont {Tantawi}, \citenamefont {Dolgashev}, \citenamefont
  {Nantista}, \citenamefont {Higashi}, \citenamefont {Aicheler}, \citenamefont
  {Heikkinen},\ and\ \citenamefont {Wuensch}}]{Exp_study_RF_pulsed_heating}%
  \BibitemOpen
  \bibfield  {author} {\bibinfo {author} {\bibfnamefont {L.}~\bibnamefont
  {Laurent}}, \bibinfo {author} {\bibfnamefont {S.}~\bibnamefont {Tantawi}},
  \bibinfo {author} {\bibfnamefont {V.}~\bibnamefont {Dolgashev}}, \bibinfo
  {author} {\bibfnamefont {C.}~\bibnamefont {Nantista}}, \bibinfo {author}
  {\bibfnamefont {Y.}~\bibnamefont {Higashi}}, \bibinfo {author} {\bibfnamefont
  {M.}~\bibnamefont {Aicheler}}, \bibinfo {author} {\bibfnamefont
  {S.}~\bibnamefont {Heikkinen}}, \ and\ \bibinfo {author} {\bibfnamefont
  {W.}~\bibnamefont {Wuensch}},\ }\href {\doibase
  10.1103/PhysRevSTAB.14.041001} {\bibfield  {journal} {\bibinfo  {journal}
  {Phys. Rev. ST Accel. Beams}\ }\textbf {\bibinfo {volume} {14}},\ \bibinfo
  {pages} {041001} (\bibinfo {year} {2011})}\BibitemShut {NoStop}%
\bibitem [{\citenamefont {Tantawi}\ \emph {et~al.}(2010)\citenamefont
  {Tantawi}, \citenamefont {Dolgashev}, \citenamefont {Higashi},\ and\
  \citenamefont {Spataro}}]{HighGradAcc_pulsehheating}%
  \BibitemOpen
  \bibfield  {author} {\bibinfo {author} {\bibfnamefont {S.~G.}\ \bibnamefont
  {Tantawi}}, \bibinfo {author} {\bibfnamefont {V.}~\bibnamefont {Dolgashev}},
  \bibinfo {author} {\bibfnamefont {Y.}~\bibnamefont {Higashi}}, \ and\
  \bibinfo {author} {\bibfnamefont {B.}~\bibnamefont {Spataro}},\ }\href
  {\doibase 10.1063/1.3520332} {\bibfield  {journal} {\bibinfo  {journal} {AIP
  Conference Proceedings}\ }\textbf {\bibinfo {volume} {1299}},\ \bibinfo
  {pages} {29} (\bibinfo {year} {2010})}\BibitemShut {NoStop}%
\bibitem [{\citenamefont {Pritzkau}\ \emph {et~al.}(1999)\citenamefont
  {Pritzkau}, \citenamefont {Bowden}, \citenamefont {Menegat},\ and\
  \citenamefont {Siemann}}]{pritzkau1999possible_pulsed_heating}%
  \BibitemOpen
  \bibfield  {author} {\bibinfo {author} {\bibfnamefont {D.~P.}\ \bibnamefont
  {Pritzkau}}, \bibinfo {author} {\bibfnamefont {G.~B.}\ \bibnamefont
  {Bowden}}, \bibinfo {author} {\bibfnamefont {A.}~\bibnamefont {Menegat}}, \
  and\ \bibinfo {author} {\bibfnamefont {R.~H.}\ \bibnamefont {Siemann}},\ }in\
  \href@noop {} {\emph {\bibinfo {booktitle} {AIP Conference Proceedings}}},\
  Vol.\ \bibinfo {volume} {474}\ (\bibinfo {organization} {AIP},\ \bibinfo
  {year} {1999})\ pp.\ \bibinfo {pages} {387--398}\BibitemShut {NoStop}%
\bibitem [{\citenamefont {Pritzkau}(2001)}]{pritzkau2001rf}%
  \BibitemOpen
  \bibfield  {author} {\bibinfo {author} {\bibfnamefont {D.~P.}\ \bibnamefont
  {Pritzkau}},\ }\href@noop {} {\emph {\bibinfo {title} {RF pulsed heating}}},\
  \bibinfo {type} {Tech. Rep.}\ (\bibinfo {year} {2001})\BibitemShut {NoStop}%
\bibitem [{\citenamefont {Grudiev}\ \emph {et~al.}(2009)\citenamefont
  {Grudiev}, \citenamefont {Calatroni},\ and\ \citenamefont {Wuensch}}]{sc}%
  \BibitemOpen
  \bibfield  {author} {\bibinfo {author} {\bibfnamefont {A.}~\bibnamefont
  {Grudiev}}, \bibinfo {author} {\bibfnamefont {S.}~\bibnamefont {Calatroni}},
  \ and\ \bibinfo {author} {\bibfnamefont {W.}~\bibnamefont {Wuensch}},\
  }\href@noop {} {\bibfield  {journal} {\bibinfo  {journal} {Physical Review
  Special Topics-Accelerators and Beams}\ }\textbf {\bibinfo {volume} {12}},\
  \bibinfo {pages} {102001} (\bibinfo {year} {2009})}\BibitemShut {NoStop}%
\bibitem [{\citenamefont {Willoughby}(2005)}]{willoughby2005elastically_avg}%
  \BibitemOpen
  \bibfield  {author} {\bibinfo {author} {\bibfnamefont {P.~P.~J.}\
  \bibnamefont {Willoughby}},\ }\emph {\bibinfo {title} {Elastically averaged
  precision alignment}},\ \href@noop {} {Ph.D. thesis},\ \bibinfo  {school}
  {Massachusetts Institute of Technology} (\bibinfo {year} {2005})\BibitemShut
  {NoStop}%
\bibitem [{\citenamefont {Borchard}\ \emph {et~al.}(2018)\citenamefont
  {Borchard}, \citenamefont {Appert},\ and\ \citenamefont
  {Hoh}}]{borchard2018fabrication}%
  \BibitemOpen
  \bibfield  {author} {\bibinfo {author} {\bibfnamefont {P.}~\bibnamefont
  {Borchard}}, \bibinfo {author} {\bibfnamefont {S.}~\bibnamefont {Appert}}, \
  and\ \bibinfo {author} {\bibfnamefont {J.}~\bibnamefont {Hoh}},\ }in\
  \href@noop {} {\emph {\bibinfo {booktitle} {Journal of Physics: Conference
  Series}}},\ Vol.\ \bibinfo {volume} {1067}\ (\bibinfo {organization} {IOP
  Publishing},\ \bibinfo {year} {2018})\ p.\ \bibinfo {pages}
  {082002}\BibitemShut {NoStop}%
\bibitem [{\citenamefont {Steele}(1966)}]{steele1966nonresonant}%
  \BibitemOpen
  \bibfield  {author} {\bibinfo {author} {\bibfnamefont {C.~W.}\ \bibnamefont
  {Steele}},\ }\href@noop {} {\bibfield  {journal} {\bibinfo  {journal} {IEEE
  Transactions on Microwave theory and Techniques}\ }\textbf {\bibinfo {volume}
  {14}},\ \bibinfo {pages} {70} (\bibinfo {year} {1966})}\BibitemShut {NoStop}%
\bibitem [{\citenamefont {Wang}(2004)}]{wang2004accelerator}%
  \BibitemOpen
  \bibfield  {author} {\bibinfo {author} {\bibfnamefont {J.}~\bibnamefont
  {Wang}},\ }\href@noop {} {\emph {\bibinfo {title} {Accelerator structure
  development for NLC/GLC}}},\ \bibinfo {type} {Tech. Rep.}\ (\bibinfo
  {institution} {Stanford Linear Accelerator Center, Menlo Park, CA (US)},\
  \bibinfo {year} {2004})\BibitemShut {NoStop}%
\bibitem [{\citenamefont {Tantawi}\ \emph {et~al.}(2005)\citenamefont
  {Tantawi}, \citenamefont {Nantista}, \citenamefont {Dolgashev}, \citenamefont
  {Pearson}, \citenamefont {Nelson}, \citenamefont {Jobe}, \citenamefont
  {Chan}, \citenamefont {Fant}, \citenamefont {Frisch},\ and\ \citenamefont
  {Atkinson}}]{Tantawi2005}%
  \BibitemOpen
  \bibfield  {author} {\bibinfo {author} {\bibfnamefont {S.~G.}\ \bibnamefont
  {Tantawi}}, \bibinfo {author} {\bibfnamefont {C.~D.}\ \bibnamefont
  {Nantista}}, \bibinfo {author} {\bibfnamefont {V.~A.}\ \bibnamefont
  {Dolgashev}}, \bibinfo {author} {\bibfnamefont {C.}~\bibnamefont {Pearson}},
  \bibinfo {author} {\bibfnamefont {J.}~\bibnamefont {Nelson}}, \bibinfo
  {author} {\bibfnamefont {K.}~\bibnamefont {Jobe}}, \bibinfo {author}
  {\bibfnamefont {J.}~\bibnamefont {Chan}}, \bibinfo {author} {\bibfnamefont
  {K.}~\bibnamefont {Fant}}, \bibinfo {author} {\bibfnamefont {J.}~\bibnamefont
  {Frisch}}, \ and\ \bibinfo {author} {\bibfnamefont {D.}~\bibnamefont
  {Atkinson}},\ }\href {\doibase 10.1103/PhysRevSTAB.8.042002} {\bibfield
  {journal} {\bibinfo  {journal} {Phys. Rev. ST Accel. Beams}\ }\textbf
  {\bibinfo {volume} {8}},\ \bibinfo {pages} {042002} (\bibinfo {year}
  {2005})}\BibitemShut {NoStop}%
\bibitem [{\citenamefont {Nasr}\ and\ \citenamefont
  {Tantawi}(2017)}]{nasr2017novel}%
  \BibitemOpen
  \bibfield  {author} {\bibinfo {author} {\bibfnamefont {M.}~\bibnamefont
  {Nasr}}\ and\ \bibinfo {author} {\bibfnamefont {S.}~\bibnamefont {Tantawi}},\
  }in\ \href@noop {} {\emph {\bibinfo {booktitle} {8th Int. Particle
  Accelerator Conf.(IPAC'17), Copenhagen, Denmark, 14{\^a} 19 May, 2017}}}\
  (\bibinfo {organization} {JACOW, Geneva, Switzerland},\ \bibinfo {year}
  {2017})\ pp.\ \bibinfo {pages} {1634--1636}\BibitemShut {NoStop}%
\bibitem [{\citenamefont {Nasr}\ and\ \citenamefont
  {Tantawi}(2018{\natexlab{b}})}]{nasr2018design_dual}%
  \BibitemOpen
  \bibfield  {author} {\bibinfo {author} {\bibfnamefont {M.}~\bibnamefont
  {Nasr}}\ and\ \bibinfo {author} {\bibfnamefont {S.}~\bibnamefont {Tantawi}},\
  }in\ \href@noop {} {\emph {\bibinfo {booktitle} {9th Int. Particle
  Accelerator Conf.(IPAC'18), Vancouver, BC, Canada, April 29-May 4, 2018}}}\
  (\bibinfo {organization} {JACOW Publishing, Geneva, Switzerland},\ \bibinfo
  {year} {2018})\ pp.\ \bibinfo {pages} {4391--4394}\BibitemShut {NoStop}%
\bibitem [{\citenamefont {Welander}\ \emph {et~al.}(2018)\citenamefont
  {Welander}, \citenamefont {Li}, \citenamefont {Nasr},\ and\ \citenamefont
  {Tantawi}}]{welander2018parallel}%
  \BibitemOpen
  \bibfield  {author} {\bibinfo {author} {\bibfnamefont {P.}~\bibnamefont
  {Welander}}, \bibinfo {author} {\bibfnamefont {Z.}~\bibnamefont {Li}},
  \bibinfo {author} {\bibfnamefont {M.}~\bibnamefont {Nasr}}, \ and\ \bibinfo
  {author} {\bibfnamefont {S.}~\bibnamefont {Tantawi}},\ }in\ \href@noop {}
  {\emph {\bibinfo {booktitle} {9th Int. Particle Accelerator Conf.(IPAC'18),
  Vancouver, BC, Canada, April 29-May 4, 2018}}}\ (\bibinfo {organization}
  {JACOW Publishing, Geneva, Switzerland},\ \bibinfo {year} {2018})\ pp.\
  \bibinfo {pages} {3835--3837}\BibitemShut {NoStop}%
\end{thebibliography}%

\end{document}